\documentclass[showpacs,prl]{revtex4}
\usepackage{amssymb,amsmath}
\newcommand{\ds}{\displaystyle}

\def\nc{\newcommand}
\def\rr{\rightarrow}
 \nc{\Etot}{{E}}
 \nc{\Eeq}{{E^{\rm eq}_1}}
 \nc{\Teq}{{T_{\rm eq}}}
 \nc{\taumix}{{\tau_{\rm mix}}}

 \def\bfX{{\bf X}}
 
 \def\Jac{{\rm Jac}}
 \def\div{{\rm div}}
 \def\calN{{\cal N}}
\begin{document}
\title{
Micro-canonical thermodynamics:
Why does heat flow from hot to cold. 
}
\author{Hans Henrik Rugh }
\affiliation{Department of Mathematics, University of Paris-Sud 11,
           FR-91405 Orsay, France}
\date{\today}
 \begin{abstract}
We show how to use a central limit approximation for additive co-cycles
to describe  non-equilibrium and far from equilibrium thermodynamic
behaviour.  We consider first two weakly coupled Hamiltonian dynamical
systems initially at different micro-canonical temperatures.
We describe a stochastic model where
the energy-transfer between the
two systems is considered as a random variable satisfying
a central limit approximation.
We show that fluctuations in energy observables 
are linearly related to the heat-transfer (dissipation).
As a result, on average, heat flows from hot to cold.
We also consider the far from equilibrium  situation
of a non-Hamiltonian thermostatted system as
in Evans et al. {\em Phys.\ Rev.\ Lett.} {\bf 71}, 2401 (1993).
Applying the same central limit approximation we re-derive
their relation for the
violation of the 2nd law of thermodynamics. 
We note that time-reversal symmetry is not used in our derivation.
 \end{abstract}
\maketitle

Starting with the pioneering work of Boltzmann,
equilibrium statistical mechanics has
developped into a solid corner-stone of theoretical physics.
Using perturbative expansions,
Kubo considered systems close to equilibrium and
obtained fluctuation-dissipation theorems,
relating dissipation (linear response) in the system to
fluctuations described by decay of correlation functions.
We refer to e.g. \cite{Kubo66} for a nice review.
Of great current interest, but less understood,
is the case of far from equilibrium statistical mechanics.
Recent approaches to the subject were initiated by 
Evans et al. \cite{ECM93} (see also \cite{ES02} for a review).
One considers a thermostated system 
driven by external forces. 
The thermostat gives rise to a phase space contraction
which is interpreted as a production of entropy. Through numerical
simulations, Evans et al. made the interesting observation that 
the 2nd law of thermodynamics is broken in a systematic way.
As a model for this phenomena the authors suggest that 
the dynamical behavior ressembles that of the attractor
of an Anosov flow  with an underlying time-reversal symmetry.
This model
has then been further developped
by e.g.\  Gallavotti and Cohen \cite{GC95}.
We also refer to Kurchan \cite{Kurchan98} and
Lebowitz and Spohn \cite{LS99} for a somewhat different stochastic approach
(but still using an inherent time-reversal symmetry). 
Bustamante et al. \cite{BLR05} (see also references therein) 
gives a review of recent physical experiments supporting the theoretical
work.

Our aim below is to provide an elementary description
of not only the above mentioned phenomena, but also the
time evolution of the energy transfer between two 
weakly coupled hamiltonian systems. Neither thermostats
nor external forces are involved in this
latter case. 
Our arguments are based upon a strong
stochastic assumption, namely that
a central limit approximation
applies to so-called additive co-cycles in the systems.
In particular, we do not need the presence of
a time-reversal symmetry.
As in \cite{ECM93}
we obtain in both cases a universal law for the violation
of the 2nd law of thermodynamics.
Our approach is closely related to the study of
the "structure functions" which were used by Zwanzig \cite{Z61}.

\section{Two systems and a  weak interaction}

Consider hamiltonians $H_1$ and $H_2$ on 
phase spaces $\Omega_1$ and $\Omega_2$, respectively.
We will study the energy flow between the
two systems that arises from introducing
a weak coupling $U_{12}$ defined on the product space,
$\Omega_1\times \Omega_2$.
When the (micro-canonical) temperatures of the two systems
are different one expects on heuristic grounds that energy
should flow from the `hotter' to the `colder' system. We wish
to quantify this phenomenon within the micro-canonical
ensemble and without introducing external forces,
not heat-baths in the problem.
Recall, that the flow $\phi^t$ associated to
the total Hamiltonian function $H=H_1+H_2+U_{12}$ 
preserves the total energy as well as the
product Liouville volume $m(d\xi)=m_1(d\xi_1)m_2(d\xi_2)$,
with $\xi=(\xi_1,\xi_2)$ being coordinates on $\Omega_1\times\Omega_2$.
In particular, also the micro-canonical measure
$\mu_\Etot(d\xi)=\delta(H(\xi)-E) m(d\xi)$ having support on the energy
surface $H(\xi)=\Etot$ is invariant under the flow.
Our goal is to study the time evolution  of e.g.\ the first system,
 $H_1$ under 
the global flow $\phi^t$ at some fixed total energy 
$\Etot$. 
We will adapt  a stochastic
point of view. The basic idea is that when the coupling
is weak and each sub-system (hopefully) is mixing fast enough
on its proper energy
surface then at every instant of time each system is close
to an equilibrium of that system.
In spirit this is close to linear response theory.
There, however, one usually applies a fixed (small)
perturbation and then lets the system evolve to a new equilibrium
state close to the original. In our context we consider a slow
but steady evolution away from the original state.
Instead of starting
out at a particular point in phase space we
start out with  
an $H_1$-conditional
ensemble at time $t=0$:
\[ \delta(H_1(\xi)-E_1) \mu_\Etot(d\xi).
\]
The normalizing factor,
\[ e^{f(E_1)}= 
 \int \delta(H_1(\xi)-E_1) \mu_\Etot(d\xi),\]
defines the $E_1$-conditional entropy, $f(E_1)$.
Now, this conditional
ensemble is, in general, not time-invariant when
the interaction is turned on.
At time $t>0$, the probability distribution
of the values of $H_1(\phi^t(\xi))$ will be
given by an expression
of the form~:

\[ p_t(x|E_1) = 
    \frac{\int 
      \delta \left( H_1(\phi^t( \xi))-x \right)
      \delta \left( H_1(\xi)-E_1 \right) \mu_\Etot(d\xi) }
    {\int 
      \delta \left( H_1(\xi)-E_1 \right) \mu_\Etot(d\xi) }
\]
We assume for simplicity of notation
that the distribution admits a density. More generally one
could formulate the relation in terms of measures
without affecting the conclusions.
By normalization, the kernel satisfies:
\[ \int p_t(x|E_1) \; dx = 1 .\]
It verifies, however, another important identity:
Multiplying by $e^{f(E_1)}$ and integrating with respect to $E_1$
we get
\begin{eqnarray}
  \int p_t(x|E_1) e^{f(E_1)} dE_1 &=&
    \int \delta \left( H_1(\phi^t (\xi))-x \right) \mu_\Etot(d\xi) 
    \nonumber \\
 &=&
    \int \delta \left( H_1(\xi)-x \right) \mu_\Etot(d\xi)
    \nonumber \\
    &=& e^{f(x)}
      \label{invar dens}
     ,
\end{eqnarray}
where we used the fact that $\mu_\Etot(d\xi)$ is $\phi^t$-invariant.
At present the above expressions are exact.

We come to the crucial approximation:
 Consider $X_t=H_1(\phi^t(\xi))$
as a random variable whose probability distribution
is given by 
$p_t(x|E_1)dx$.
The mean drift is $m_t(E_1)= {\mathbf E}(X_t)-E_1=
\int (x-E_1) p_t(x|E_1) \; dx$
and the variance is $\sigma_t^2(E_1) = {\rm Var}(X_t)$.
Both drift and variance are functions of $t$ and $E_1$.
When $t$ tends to zero, 
$p_t(x|E_1) \rr \delta(x-E_1)$ and
when $t$ tends to infinity (assuming global mixing)
$p_t(x|E_1) \rr e^{f(x)} \times {\rm const}$.
The energy increment
$A^t(\xi)=H_1(\phi^t(\xi))-H_1(\xi)$ 
is an `additive co-cycle'. By this we mean  that 
for all $s,t\geq 0$, $\xi\in \Omega_1\times\Omega_2$:
\begin{equation}
A^{t+s}(\xi) = A^t \circ \phi^s(\xi) +A^s(\xi).
\label{add cocycle}
\end{equation}

When time-correlations decay
fast enough, such additive co-cycles
tend to have asymptotic properties in common with sums of independent
random variables. In particular, it may be within reason to 
assume that it behaves like a gaussian variable
(at least on certain
time scales). This is the case e.g.\ when looking at
smooth observables in Anosov systems
or in exponentially mixing Markov chains.\\

Central Limit Approximation:
Assume that there
are two characteristic time-scales $\taumix \ll \Teq$ where
$\taumix$ is a `mixing'-time of the sub-systems and
$\Teq$ is a time-scale for `significant' changes in the
energy of each sub-systems. 
When $\taumix\ll t \ll \Teq$ 
we may approximate $p_t(x|E_1)$ 
by the corresponding normal distribution:
\begin{equation}
p_t(x|E_1)  \approx \frac{1}{\sqrt{2\pi} \sigma_t}
    \exp \left( - \frac{(x-E_1-m_t)^2}{2 \sigma_t^2} \right).
     \label{central limit}
\end{equation}
Weak Coupling Approximation: We need to be able to calculate
derivatives of 
the relative entropy $f(x)$.  This is easy if we can 
neglect the interaction term.
In this approximation $H(\xi)=H_1(\xi_1)+H_2(\xi_2)$
and
\begin{eqnarray}
  e^{f(x)}
  &=& \int \delta \left( H_1(\xi_1)-x \right) \mu_\Etot(d\xi) \nonumber\\
  &=& \int
       \delta \left( H_1(\xi_1)-x \right) 
       \delta \left( H_1(\xi_2)-(E-x) \right)
       m_1(d\xi_1)
       m_2(d\xi_2)
       \nonumber\\
  &=& e^{S_1(x) + S_2(E-x)},
 \end{eqnarray}
where 
$S_i(x)=\log \int \delta \left( H_i(\xi_i)-x \right) m_i(d\xi_i)$, $i=1,2$
       are the $\mu$-canonical entropies  of the two sub-systems.
For each sub-system,
one may associate its $\mu$-canonical 
temperature, i.e.\
$\ds \frac{1}{T_1} = \frac{\partial S_1}{\partial E_1}$ as well as
its heat-capacity
$\ds \frac{1}{C_1} =
\frac{\partial T_1}{\partial E_1}=
-T_1^2\frac{\partial^2 S_1}{\partial E_1^2}$.
As we showed in earlier papers 
\cite{Rugh97,Rugh98}, such quantities are computable within the 
$\mu$-canonical ensemble provided each subsystem is ergodic. 
On the time-scale 
$\taumix\ll t \ll \Teq$,
one may assign local micro-canonical
thermodynamic characteristics to each sub-system.
In the weak coupling limit
the global equilibrium of the combined system
is at the energy $E_1=\Eeq$ for which
$f'(\Eeq)=0$ or $T_1(\Eeq)=T_2(\Etot-\Eeq)$, i.e.\ the two micro-canonical
temperatures are equal. When heat-capacities are positive
(i.e.\ $S_1$ and $S_2$ are strictly concave) 
the corresponding energy is unique.
But we do not need this for the present discussion.

Recall that the distribution $p_t$ function for $X_t$
leaves the micro-canonical ensemble
invariant:
 \[ \int p_t(x | E_1) e^{f(E_1)} dE_1 = e^{f(x)}.\]
When $t$ is not too large it is reasonable to expect 
the variance of $X_t$ to be small compared to the inverse of
the curvature of $f$. We may then replace $f(E_1)$ by its first
order Taylor expansion around $x$:
$f(E_1)-f(x) = \lambda (E_1-x) +o(E_1-x)$ with
$\ds \lambda = f'(x) =
            \frac{1}{T_1(x)} - \frac{1}{T_2(\Etot - x)}$.
We insert this
our central limit expression (\ref{central limit}) and get
for the integral:
\[ \int
    \exp \left( - \frac{(x-E_1-m_t)^2}{2 \sigma_t^2}
    + \lambda (E_1-x) \right) dE_1 = 
      \exp \left( \lambda^2 \sigma_t^2/2 - \lambda m_t \right) = 1. \]
This implies that either $\lambda=0$ (which corresponds
to the systems having identical temperatures, i.e\ 
they are in thermodynamic
equilibrium) or, more interestingly,
when $\lambda$ is non-zero we get the relation
$m_t= \frac12 \lambda \sigma_t^2$. We have obtained the following:\\

\noindent {\bf Fluctuation Dissipation Relation.}
Under the Central Limit Approximation we have for $\taumix\ll t \ll \Teq$:
\\[-5mm]
   \begin{equation}
	 m_t(E_1) = \frac12 {\sigma_t^2(E_1)}
          \left( 
            \frac{1}{T_1(E_1)} - \frac{1}{T_2(\Etot - E_1)} \right).
	    \label{eq fluctuations}
   \end{equation}
  \begin{equation}
      \log \frac{p_t(E_1+u \;|\; E_1)}
       {p_t(E_1-u \;|\; E_1)} =  \frac{2 m_t}{\sigma_t^2} u =
          \left( 
            \frac{1}{T_1(E_1)} - \frac{1}{T_2(\Etot-E_1)} \right)  u .
      \label{eq violated law}
  \end{equation}

The first equality states that the mean drift in energy 
(dissipation) of each sub-system is proportional to the fluctuations
in the sub-system
with a constant of proportionality being the difference of 
the inverse temperatures of the two systems.
Since fluctuations are non-negative, on average energy flows
from `hot' to `cold'.
The second equality
is obtained by combining expression (\ref{central limit})
and the fluctuation dissipation relation (\ref{eq fluctuations}).
It expresses the relative probability
of a violation of the 2nd law of thermodynamics at certain time
and energy scales.

\section{Validity and computability}
Estimates both for $p_t(x|E_1)$ and temperatures
may be obtained through numerical simulations
thus allowing for a verification of our Central Limit Approximation (CLA)
as well as the Fluctuation Dissipation Relation (FDR).
Assuming ergodicity
it suffices for $p_t(x|E_1)$ to run the system without interactions
to get initial points representing the $H_1$-conditional ensemble,
then turn on the (weak) interaction and run the ensemble
to give estimates for this
transition probability.
For the temperatures of the sub-systems, we may e.g.\ 
use \cite{Rugh97,Rugh98} : 
If $\bfX_i$, $i=1,2$ are vector fields on
$T\Omega_i$ for which
$dH_i(\bfX_i)\equiv 1$ then,  without interactions, we have
$1/T_i=\langle {\rm div}_{m_i}(\bfX_i)|E_i\rangle$, the  
ergodic average of the observable $ {\rm div}_{m_i}(\bfX_i)$
at energy $E_i$ for each sub-system.
Under weak interactions but
assuming that sub-system energies varies slowly compared to 
its ergodic averaging,   such expressions are good candidates
for temperature observables for the two sub-systems.

For numerical reasons,
interactions should be small but not too small.
The interaction constitute background fluctuations of 
order $\omega={\cal O}(U_{12})$. If this is too large, it 
is unlikely that
one may observe the FD-relation.
On the other hand energy exchange is a second
order phenomena (see e.g.\  \cite{Z61}) so we have to wait
a time of order $1/\omega$ 
to get an effective energy transfer exceeding the
 background noise. Numerical errors could then create problems.

The expression (\ref{central limit}) for $p_t(x|E_1)$
could be a good approximation even on time scales
smaller than
the mixing time of sub-systems. The reason is that for a large system
space mixing may give rise to a good central limit approximation
for the observable $H_1$ even without time-mixing.
The FDR may, however, fail  in that case.
As a (non-generic, though) example suppose that system 2
itself is a sum of two sub-systems with a small interaction:
$H_2=H_{a} + V_{ab} + H_{b}$. We now add an interaction
of the form \ 
$U_{12}=U_{1,a}-V_{ab}$ to $H_1+H_2$. This has the effect
of coupling $H_1$ and $H_{a}$ but decoupling their sum from $H_{b}$.
As a result there will be fluctuations in $H_1$ but the predicted
mean drift will in general be wrong,
there will be no global equilibration and
the fluctuation dissipation relation as stated should fail.

\section{A thermostated non-equilibrium system}
We now consider a
situation as described in \cite{ECM93} in which the authors
consider a Hamiltonian $H$ on a phase space $\Omega$
and subjected to a thermostat. One associates to this a
non-Hamiltonian vector field $X_H$ and its flow $\phi^t$.
For details of the construction see \cite{ECM93}.
For our purposes, the essential properties may be summarized
as follows: The Hamiltonian flow $\phi^t$ preserves $H$ but not the 
Liouville space volume $m$. 
We write $\mu_E=\delta(H-E) m$ for the  associated
area form on the energy surface.
Neither $m$, nor $\mu_E$ is 
invariant under $X_H$.
The Jacobian 
$\Jac^t(\xi) = m(\phi^t d\xi)/m(d \xi)$ describes the
volume transformation along the flow.
Because of $H$-invariance  we have
$\Jac^t(\xi)= \mu_E(\phi^t dA)/\mu_E(d A)$ as well,
i.e.\ it is the same Jacobian
for the surface area and for the bulk volume.
To see this one may e.g.\ use differential
forms and take the Lie derivative of $\mu_H$:
$L_{X_H} \left(\delta(H-E) m\right) = \delta(H-E) L_{X_H} m=
\delta(H-E) 
\div_m(X_H)
m$, where the first equality is due to $L_X H=dH(X)=0$.
It shows that infinitesimally 
$m$ and $\mu_E$ have the same Jacobian so this is also the case
for the flow.

One wants to observe
the phase space contraction rate,
manifested by the above Jacobian. 
The Jacobian is multiplicative and not additive but
taking a logarithm, we get an additive co-cycle as before.
So, our observable will be 
$A^t=\log \Jac^t(\xi)$, which 
verifies
$A^{t+s}= A^t \circ \phi^s + A^s$. 
It is a computable, i.e.\ observable quantity in this context.
The object of interest is then the distribution function for 
$A^t$ which 
we consider in the  $\mu_E$-ensemble:
\[ p_t(\alpha) = \frac{\int_\Omega \delta(A^t(\xi)-\alpha) \mu_E }
      { \int_\Omega \mu_E}.\]
Since $\phi^t : \Omega \rr \Omega$ is a diffeomorphism we get
by change of variables:
\[
\int_{\Omega} \mu_E =
\int_{\phi^t \Omega} \mu_E =
\int_{\Omega} \Jac^t(\xi) \; \mu_E(\xi) .\]
Inserting the distribution function for $A^t$ we get the (exact) relation:
\[ 1 = \frac
{ \int_{\Omega} \Jac^t(\xi) \; \mu_E(\xi)}
{ \int_{\Omega}  \mu_E(\xi)} = \int e^\alpha p_t (\alpha) \; d\alpha.
   \]
This is a constraint equation for the distribution function $p_t$.
As $A_t$ is an additive co-cycle we again make
the strong stochastic assumption
that we may approximate $p_t$ by a normal distribution,
$p_t \sim \calN (m_t,\sigma_t^2)$. Doing so and inserting 
in the above
constraint equation yields $\; \exp(m_t+\sigma_t^2/2)=1\; $ or
$m_t=-\sigma_t^2/2$.
This is the FD relation of e.g.\  \cite{ECM93,GC95}). And
as in these cited papers
one has the symmetry-relation
 \[ p_t(\alpha)/p_t(-\alpha) = e^\alpha.\]

 There is an important 
approximation taking place when comparing
the above derivation and the numerical simulations.
In our derivation the distribution of the observable
$A_t$ is with respect to the initial distribution
$\mu_E=\delta(H-E)m$ whereas the numerical computations are 
done for a (hopefully) stationary state
of the system. 
This distinction also
makes a subtle difference in the point of view
of \cite{ECM93} and \cite{GC95}.
For an Anosov system, the distinction
is not important. In both ensembles a central limit approximation
hold and with the same constants. Working in the stationary
state, however, is numerically more stable as it
eliminates the contribution of transients which
can be quite large.
 For more realistic models,
 it would be of interest to compare numerically
the two ensemble distributions.

We note that time-reversal symmetry
is not needed in the above derivation. 
Again it would be interesting to compare with
numerical simulations for a system without
time-reversal symmetry. Mittag et al presented such a system
in \cite{MEW06} for which the distribution function $p_t$ was
quite far from gaussian and the FD-relation fails. This, however, 
does not contradict the above derivation since in their case
the external field was changed during the time span
of the experiment. $A^t$ is then not an additive co-cycle
so transient behaviour becomes significant.
We also note that our derivation does not make use of the
underlying sympltic structure of phase space. So in principle
our derivation makes sense for any dynamical system on
a compact manifold (here, $\{H=E\}$) that converges fast enough
towards a natural measure and for which
 correlations decay fast enough.

\end{document}